\documentclass[twocolumn,aps,prd,showpacs,nofootinbib,superscriptaddress,preprintnumbers,floatfix,10pt]{revtex4-1}

\usepackage[english]{babel} 
\usepackage[utf8]{inputenc}
\usepackage[T1]{fontenc}
\usepackage{lmodern}

\usepackage{amsmath}
\usepackage{amsfonts}
\usepackage{amssymb}
\usepackage{slashed}

\usepackage{epsfig}
\usepackage{graphicx}
\usepackage{color}
\usepackage{epstopdf}

\usepackage{hyperref}

\newcommand{\I}{\mathrm{i}}
\newcommand{\E}{\mathrm{e}}

\newcommand{\STr}{\text{STr}}

\newcommand{\Nf}{N_{\mathrm{f}}}
\newcommand{\pt}{\partial_t}
\newcommand{\psib}{\bar{\psi}}

\newcommand{\rmB}{\mathrm{B}}
\newcommand{\rmF}{\mathrm{F}}
\newcommand{\rmt}{\mathrm{t}}
\newcommand{\htop}{h_{\mathrm{t}}}
\newcommand{\rhot}{\tilde\rho}

\newcommand{\lamportal}{\lambda_{h\psi}}

\newcommand{\Nu}{N_{\mathrm{u}}}
\newcommand{\Nh}{N_{\mathrm{h}}}

\begin{document}

\author{Aaron Held}
\email{a.held@thphys.uni-heidelberg.de}
\affiliation{Institut f\"ur Theoretische Physik, Universit\"at Heidelberg, Philosophenweg 16, 69120 Heidelberg, Germany}

\author{Ren\'{e} Sondenheimer}
\email{rene.sondenheimer@uni-jena.de}
\affiliation{Theoretisch-Physikalisches Institut, Friedrich-Schiller-Universit\"at Jena, Max-Wien-Platz 1, D-07743 Jena, Germany}

\title{Higgs stability-bound and fermionic dark matter}

\begin{abstract} 
Higgs-portal interactions of fermionic dark matter -- in contrast to fermions coupled via Yukawa interactions -- can have a stabilizing effect on the standard-model Higgs potential.
A non-perturbative renormalization-group analysis reveals that, similar to higher-order operators in the Higgs potential itself, the fermionic portal coupling can increase the metastability scale by only about one order of magnitude. 
Furthermore, this regime of very weakly coupled dark matter is in conflict with relic-density constraints.
Conversely, fermionic dark matter with the right relic abundance requires either a low cutoff scale of the effective field theory or a strongly interacting scalar sector.
This results in a triviality problem in the scalar sector which persists at the non-perturbative level. 
The corresponding breakdown of the effective field theory suggests a larger dark sector to be present not too far above the dark-fermion mass-scale.
\end{abstract}

\maketitle

\section{The Higgs potential and new physics}
The most surprising result obtained at the Large Hadron Collider might be, that the masses and couplings conspire to render the standard model consistent up to very high energy scales, possibly even all the way up to the Planck scale \cite{Bezrukov:2012sa, Buttazzo:2013uya,Gabrielli:2013hma,Bednyakov:2015sca,Iacobellis:2016eof}. It is a delicate balance between the running of gauge couplings, Yukawa couplings and the Higgs potential that results in this unexpected experimental result. Although this so-called ``desert'' disfavors new degrees of freedom to be discovered any time soon, it should maybe not be taken as discouragement. Experimental evidence suggests that our knowledge of the Universe is more complete than we might have hoped for. The vast other parts of parameter space, in which the standard model as an effective field theory (EFT) could in principle be realized, results either in an instability of the scalar potential or in a sub-Planckian triviality problem. Both of these could be interpreted as a clear sign for a breakdown of the standard model and an associated new physics scale.

Focusing on the central measurement values \cite{Patrignani:2016xqp}, the standard model as an EFT is potentially valid up to $\Lambda_\mathrm{EFT}^{(\mathrm{SM})}\approx10^{41}$ GeV. At this scale, the theory reaches a $U(1)$ Landau-pole in perturbation theory \cite{GellMann:1954fq}, reflected in a triviality problem at the non-perturbative level \cite{Gockeler:1997dn, Gockeler:1997kt, Gies:2004hy}. This would suggest that the next energy scale that theorists should be concerned about is the Planck scale, at which a joint theory of gravity and matter becomes necessary, see e.g., \cite{Shaposhnikov:2009pv, Harst:2011zx, Eichhorn:2017ylw, Eichhorn:2017lry, Eichhorn:2017als, Eichhorn:2017egq, Eichhorn:2018whv}.
On the other hand, the standard model Higgs-potential, assuming that the current central values of measured mass and coupling parameters were exact, develops a metastability. The energy scale associated with this metastability is $\Lambda_\mathrm{meta}^{(\mathrm{SM})}\approx10^{10}-10^{11}$ GeV
\cite{Bezrukov:2012sa, Buttazzo:2013uya,Gabrielli:2013hma,Bednyakov:2015sca,Iacobellis:2016eof,Eichhorn:2015kea}.
Much interesting physics can be associated with this metastability \cite{Espinosa:2017sgp,Espinosa:2018mfn,Espinosa:2018eve,Espinosa:2018euj,Markkanen:2018pdo}, and the metastable electroweak vacuum could be long-lived enough to allow for the current age of the universe \cite{Flores:1982rv,Turner:1982xj,Espinosa:1995se,Isidori:2001bm,Chigusa:2017dux,Chigusa:2018uuj}. Here, we will take the most conservative viewpoint and interpret the metastability as another sign of a breakdown of the effective theory and as a demand for new degrees of freedom below $\Lambda_\mathrm{meta}^{(\mathrm{SM})}$. 
We will therefore combine the minimum of both of these scales into the scale at which new degrees of freedom are to be expected, i.e., $\Lambda_\text{new-phys}^\text{(SM)} = \text{Min}(\Lambda_\text{meta}^\text{(SM)},\Lambda_\text{EFT}^\text{(SM)})$.

Since the standard model looks surprisingly complete, it is desirable that necessary new physics, such as dark matter, will not destroy this delicate balance.
It is well-known that an additional weakly-interacting scalar field, coupled via a Higgs portal, could act as dark matter and at the same time might fully remove a sub-Planckian metastability scale \cite{Silveira:1985rk,McDonald:1993ex,Burgess:2000yq,Bento:2001yk,Patt:2006fw,Eichhorn:2014qka}. Such scalar dark-matter models could account for the total relic density if the perturbatively renormalizable Higgs-portal coupling is sufficiently weak \cite{McDonald:1993ex,Bento:2000ah,Bento:2001yk,McDonald:2001vt,Barger:2008jx}.

In this paper, we analyze whether fermionic dark matter coupled through a Higgs portal could -- despite its perturbative non-renormalizability -- delay the metastability scale in a similar manner. Na\"ively, one might object that top-quark fluctuations are known to be the cause of the metastability in the first place: consequently one sometimes reads that fermionic fluctuations destabilize the Higgs potential. 
We will argue that the declining of the Higgs quartic coupling towards the UV is only true for fermions coupled via a Yukawa interaction $yH\bar{t}t$.
It has recently been suggested \cite{Eichhorn:2018vah}, that the fundamentally different vertex structure of fermions coupled via a Higgs-portal coupling of the form 
\begin{align}
    \mathcal{L}_\text{Higgs-portal}\sim\bar\lambda_{h\psi}\,H^\dagger H \bar\psi\psi
\end{align} 
could lead to a regime in which dark fermions, with sufficiently light mass $\bar{m}_\psi$, in fact delay the onset of the Higgs-instability scale.
This approach is different from other scenarios discussed in the literature with additional right-handed neutrinos where the stabilizing effect was caused by UV boundary conditions governed by high-scale supersymmetry \cite{Mummidi:2018nph}.

We will address these questions in the framework of EFT by use of the functional renormalization group (RG) \cite{Wetterich:1992yh}, for reviews see \cite{Berges:2000ew, Aoki:2000wm, Polonyi:2001se, Pawlowski:2005xe, Gies:2006wv, Delamotte:2007pf, Rosten:2010vm, Braun:2011pp}. 
Validity of the standard EFT requires that all dimensionless combinations, i.e., $\bar\lambda_{h\psi}\,\bar{m}_\psi$ and $\lambda_2$, the quartic Higgs self-coupling, entering cross sections remain perturbative at all energy scales $k$ below the EFT cutoff scale $\Lambda_\mathrm{EFT}$.
We will find that the following intuition persists in the non-perturbative analysis:
\begin{itemize}
    \item 
    Decoupling regime: $\bar{m}_\psi \gg \Lambda_\text{meta}^\text{(SM)}$.
    It is obvious that very heavy dark fermions will decouple and not influence the metastability scale.
    \item
    Destabilizing regime: $\bar{m}_\psi \approx \Lambda_\text{meta}^\text{(SM)}$.
    Dark fermions of a mass comparable with the metastability scale will further destabilize the Higgs sector.
    \item 
    Stabilizing regime:
    $\bar{m}_\psi \ll \Lambda_\text{meta}^\text{(SM)}$.
    Comparatively light dark matter can stabilize the Higgs-potential and increase the metastability scale. However, the non-perturbative analysis shows that portal couplings $\bar\lambda_{h\psi}>1/\Lambda_\text{meta}^\text{(SM)}$ will introduce novel Landau-pole like instabilities requiring new physics even below the original metastability scale.
\end{itemize}
In principle, the functional RG allows the analysis to be extended into the non-perturbative regime, which we define as $\bar\lambda_{h\psi}\,\bar{m}_\psi > 4\pi$. To remain as conservative as possible, we refrain from such an extension and interpret the onset of non-perturbativity as a breakdown of the conventional EFT description. This is corroborated by the non-perturbative analysis which shows that the flow equations in our approximation become singular in this regime. 
\\

In Sec.~\ref{sec:FlowEquations} we analyze the three regimes in functional RG flows in a simplified Higgs-Yukawa toy model as in \cite{Borchardt:2016xju}. 
We compute the effective potential in different approximations and validate the usefulness of the perturbative beta functions to obtain a first estimate for a possible scale of new physics.
In Sec.~\ref{sec:Pheno}, we improve our analysis to obtain reliable results for the actual metastability scale of the standard model and investigate the phenomenological implications of fermionic dark matter. 
In particular, we discuss our results in the context of the dark-matter relic density.
Assuming that the fermionic dark matter is thermally produced, the relic-density constraint will rule out a shift of the EFT cutoff scale beyond the standard-model instability scale. Thus, a larger dark sector is required, if fermionic dark matter coupled via a Higgs portal is realized in Nature.

\section{RG flow of a simple Higgs-Yukawa model with fermionic dark matter}
\label{sec:FlowEquations}
To obtain an understanding of the impact of the higher-dimensional portal coupling on the RG running of the Higgs potential, we restrict the discussion to the degrees of freedom relevant for the lower Higgs-mass bound. Therefore, we consider a real scalar field $\phi$ coupled to two different species of fermions. The first fermion species $t$ is coupled via an ordinary Yukawa coupling to the scalar field $\bar{y}_{\rmt}\,\phi\,\bar{t}\,t$ and thus represents the top quark of the standard model. The second species denoted by $\psi$ describes a fermionic dark matter particle which is a singlet with respect to the standard model gauge group and couples through a dimension-$5$ operator to the scalar field. Thus, the classical action for this toy model reads, 
\begin{align}
 S &= \int_{x} \bigg[ \frac{1}{2} \partial_{\mu}\phi\, \partial^{\mu}\phi + \frac{\bar{m}_{\phi}^{2}}{2}\phi^{2} + \frac{\bar{\lambda}_{2}}{8}\phi^{4} +  \bar{t}\,\I \slashed{\partial}\,t + \frac{\I \bar{y}_{\rmt}}{\sqrt 2} \phi\, \bar{t}\,t  \notag\\
 &\qquad\quad + \psib \,\I \slashed{\partial}\,\psi + \I \bar{m}_{\psi} \psib\,\psi + \frac{\I \bar{\lambda}_{h\psi}}{2} \phi^{2} \psib\,\psi \bigg],
 \label{eq:action}
\end{align}
where we work in Euclidean signature. The couplings and mass parameters with an overbar indicate that these are dimensionful quantities.
This model strongly simplifies the electroweak sector. Nevertheless, it contains the main ingredients to discuss the generic properties of the lower Higgs-mass bound and the stability properties of the effective Higgs potential \cite{Holland:2003jr,Branchina:2005tu,Gies:2013fua,Gies:2017ajd}. Note that the model under consideration is invariant under a discrete chiral symmetry $\phi \to - \phi$, $t \to \E^{\I\frac{\pi}{2}{\gamma_{5}}}t$, $\bar{t} \to \bar{t}\, \E^{\I\frac{\pi}{2}{\gamma_{5}}}$ as well as a continuous $\mathrm{U}(1)$ symmetry $\psi \to \E^{\I\alpha}\psi$, $\bar{\psi} \to \E^{-\I\alpha}\bar{\psi}$. 
$\gamma_{5}$ is the product of the Dirac matrices as usual.
The discrete chiral symmetry mimics the electroweak symmetry group of the standard model and forbids the occurrence of a mass term for the top quark.

Investigating the RG flow of the model, we are confronted with the perturbative nonrenormalizability of the dimension-$5$ coupling $\bar{\lambda}_{h\psi}$. It was pointed out in Ref.~\cite{Eichhorn:2018vah} that also the running and threshold effects of the dark fermion mass term $\bar{m}_{\psi}$ might play an important role in the running of the Higgs sector. Hence, our investigation has to go beyond conventional analyses in the deep Euclidean region of the RG flow.

A well-suited tool capable of handling all these tasks is the functional RG. In its modern form, the running of a scale-dependent effective average action $\Gamma_{k}$ is studied. The effective average action interpolates smoothly between the classical action at a high UV cutoff scale $\Gamma_{\Lambda} = S$ and the full quantum effective action $\Gamma_{k=0} = \Gamma$, the generating functional of $1$PI Greens functions. The RG-scale dependence of $\Gamma_{k}$ is governed by the functional RG equation \cite{Wetterich:1992yh}, cf.~also \cite{Morris:1993qb}
\begin{align}
    \label{eq:wetterichEq}
 \pt \Gamma_{k} = \frac{1}{2} \STr \left[ \big(\Gamma_{k}^{(2)} + R_{k}\big)^{-1} \pt R_{k} \right]
\end{align}
which has a technically simple one-loop structure but contains the full propagator at each RG step. The regulator function $R_{k}$ ensures IR and UV finiteness. Its presence in the denominator, where it acts like a scale-dependent mass term, regulates IR divergences. Its scale-derivative in the numerator regulates UV divergences by implementing   Wilson's idea to integrate out fluctuations in a momentum-shell wise fashion. 
We refer to the literature for more details on the functional RG \cite{Berges:2000ew,Delamotte:2007pf,Gies:2006wv,Pawlowski:2005xe,Polonyi:2001se,Rosten:2010vm}.

In order to investigate the running of the Higgs potential, the Yukawa coupling, and the portal coupling, we choose the following ansatz for the effective average action based on a systematic derivative expansion,
\begin{align}
 \Gamma_{k} &= \int_{x} \bigg[ \frac{Z_{\phi}}{2} \partial_{\mu}\phi \partial^{\mu}\phi + U(\rho) + Z_{\rmt}\, \bar{t} \I \slashed{\partial} t + \frac{\I}{\sqrt{2}} H_{\mathrm{t}}(\rho) \phi\, \bar{t}t  \notag\\
 &\qquad\quad + Z_{\psi}\, \psib \I \slashed{\partial}\psi + \I H_{\psi}(\rho) \psib\psi \bigg].
\end{align}
We have introduced a generic potential $U(\rho)$ for the scalar self-interactions. 
Further we promote the Yukawa coupling as well as the dark-matter coupling to functions of the symmetry invariant $\rho = \phi^{2}/2$, $H_{\mathrm{t}}(\rho)$ and $H_{\psi}(\rho)$, respectively. 
Another advantage of the functional RG setup is that the running of full coupling functions depending on various mass scales can be studied. In particular, these beta-functionals serve as closed-form expressions to extract the flow equations for any higher-order coupling in terms of the scalar field by suitable expansions of the various potentials in polynomials.

It is convenient and useful to introduce renormalized fields to fix the RG invariance of field rescalings and to introduce dimensionless quantities, i.e.,
\begin{align}
 \rho &= Z_{\phi} k^{d-2} \tilde\rho, \qquad\qquad  &u &= k^{-d}U, \notag\\ 
 \htop &= Z_{\phi}^{-\frac{1}{2}}Z_{\psi}^{-1} k^{\frac{4-d}{2}} H_{\mathrm{t}},  &h_{\psi} &= Z_{\psi}^{-1}k^{-1}H_{\psi}.
\end{align}
This allows to track the flow over many orders of magnitude of the RG scale.
Finally, the flow equations for the potentials can be derived from Eq.~\eqref{eq:wetterichEq}. In terms of dimensionless, renormalized variables they read
\begin{align}
 \pt u(\rhot)  &=  - d\,u + \left( d-  2 + \eta_\phi \right)\rhot u'  + 2v_{d}\, \Big[ l_0^{(B)d}(\omega_\phi; \eta_{\phi}) \notag \\
 &\quad - d_\gamma\, l_0^{(\rmF)d}(\omega_\rmt; \eta_{\rmt}) -  d_\gamma \Nf\, l_0^{(\rmF)d}(\omega_\psi; \eta_{\psi}) \Big],
 \label{eq:flowU} \\
 \pt \htop(\rhot) &= \frac{1}{2} (d - 4+ \eta_{\phi} + 2\eta_{\rmt})\htop  +  (d - 2 + \eta_{\phi}) \rhot \htop'  \notag\\
 &\quad + 2 v_{d}\,  \htop (\htop + 2\rhot \htop')^{2}\,  l_{1,1}^{(\rmB\rmF)d}(\omega_{\phi},\omega_{\rmt}; \eta_{\phi},\eta_{\rmt})  \notag\\
 &\quad - 2 v_{d}\, (3\htop' + 2\rhot \htop'')\, l_{1}^{(\rmB)d}(\omega_{\phi};\eta_{\phi}), 
 \label{eq:flowHt} \\
 \pt h_{\psi}(\rhot) &= (-1 + \eta_{\psi})h_{\psi}  +  (d - 2 + \eta_{\phi}) \rhot h_{\psi}'  \notag\\
 &\quad + 8 v_{d}\,  \rhot\, h_{\psi} {h'}^{2}_{\!\!\psi}\,  l_{1,1}^{(\rmB\rmF)d}(\omega_{\phi},\omega_{\psi}; \eta_{\phi},\eta_{\psi})  \notag\\
 &\quad - 2 v_{d}\, (h_{\psi}' + 2\rhot\, h_{\psi}'')\, l_{1}^{(\rmB)d}(\omega_{\phi};\eta_{\phi}). 
 \label{eq:flowHpsi}
\end{align}
Here, primes denote derivatives with respect to $\rhot$ and $\omega_{\phi} = u' + 2\rhot u''$, $\omega_{\rmt} = \rhot \htop^{2}$, and $\omega_{\psi} = h_{\psi}^{2}$ are RG-scale and field-amplitude dependent mass parameters for the different fields. Further, $d$ denotes the spacetime dimension, $d_{\gamma}$ is the dimension of the Clifford algebra and $v_{d}^{-1} = 2^{d+1}\pi^{\frac{d}{2}}\Gamma(d/2)$. The running of the wave function renormalizations $Z_{\phi/\rmt/\psi}$ is encoded in the anomalous dimensions of the fields, $\eta_{\phi/\rmt/\psi} = - \pt \ln  Z_{\phi/\rmt/\psi}$. Their flow is governed by
\begin{align}
    \label{eq:etaPhi}
 \eta_{\phi} &=  \frac{4 v_d}{d} \Big\{ 2\kappa  \big(3 u^{\prime\prime} + 2\kappa u^{\prime\prime\prime}\big)^2 \, m_2^{(\rmB)d}(\omega_{\phi};\eta_\phi) \notag \\
    &\quad +  d_\gamma (\htop + 2\kappa\htop')^2 \Big[ m_4^{(\rmF)d}(\omega_\rmt;\eta_\rmt)  - \kappa \htop^2   m_2^{(\rmF) d} (\omega_\rmt;\eta_\rmt) \Big]   \notag \\
     &\quad + 4 d_{\gamma}\kappa h_{\psi}' \Big[ m_4^{(\rmF)d}(\omega_\psi;\eta_\psi)  -  h_{\psi}^2   m_2^{(\rmF) d} (\omega_\psi;\eta_\psi) \Big] \Big\}_{\rhot = \kappa},\\
 \eta_{\rmt} &= \frac{4 v_d}{d} (\htop+2\kappa \htop')^{2} m_{1,2}^{(\rmF\rmB)d}(\omega_\rmt,\omega_{\phi};\eta_\rmt,\eta_\phi) \Big|_{\rhot = \kappa},\\
 \eta_{\psi} &= \frac{16 v_d}{d} \kappa\, h_{\psi}^{\prime 2} \, m_{1,2}^{(\rmF\rmB)d}(\omega_{\psi},\omega_{\phi};\eta_{\psi},\eta_{\phi})\Big|_{\rhot = \kappa},
\end{align}
where the right-hand sides are evaluated at $\rhot = \kappa$. The latter is the actual scale-dependent minimum of the scalar potential. It lies at vanishing field amplitude in the symmetric (SYM) regime and at nonvanishing field values in case of spontaneous symmetry breaking (SSB). 
The so-called threshold functions $l$ and $m$ on the right-hand sides of the flow equations arise from the integration over loop momenta and contain the nonuniversal regulator dependence. They can be evaluated analytically for a piece-wise linear regulator function which is optimized with respect to the present truncation scheme \cite{Litim:2000ci,Litim:2001up} and can be found, e.g., in \cite{Gies:2017zwf}.

The non-perturbative RG evolution of Yukawa models has been investigated extensively in many contexts in the literature. This includes effective quark-meson models \cite{Jungnickel:1995fp,Bohr:2000gp,Braun:2008pi,Pawlowski:2014zaa,Braun:2014ata,Rennecke:2016tkm}, statistical systems \cite{Rosa:2000ju,Hofling:2002hj,Braun:2010tt,Diehl:2007th,Classen:2015mar}, as well as asymptotic freedom and safety \cite{Vacca:2015nta,Eichhorn:2018vah,Gies:2018vwk}.
In particular, a simple Higgs-Yukawa toy model has proven useful to investigate the generic mechanisms and properties for the occurrence of lower Higgs-mass bounds within the set of power-counting renormalizable operators and beyond \cite{Holland:2003jr,Branchina:2005tu,Fodor:2007fn,Branchina:2008pc,Gies:2013fua,Eichhorn:2015kea,Borchardt:2016xju,Gies:2017zwf,Sondenheimer:2017jin}.

\subsection{Preliminary analysis of the quartic polynomial potential}

The basic mechanisms  at work can already be analyzed in polynomial approximations of the potentials. In the symmetric regime, we expand the three different potentials in power series around vanishing field amplitude, 
\begin{align}
 u(\rhot) &= \sum_{n=1}^{\Nu} \frac{\lambda_n}{n!}\rhot^{n}, \qquad \htop(\rhot) = \sum_{n=0}^{\Nh} \frac{y_n}{n!}\rhot^{n}, \notag \\ 
 h_{\psi}(\rhot) &= \sum_{n=0}^{N_{\psi}} \frac{\lambda_{h\psi,n}}{n!}\rhot^{n}.
\end{align}
In this parametrization, the coefficients have the natural interpretation of conventional coupling constants and mass terms. For instance, the dimensionless scalar mass-parameter is given by the linear term of the scalar potential $u$, i.e., $\lambda_{1} \equiv m_{\phi}^{2}$. The quartic coupling is given by $\lambda_{2}$. The higher-order coefficients describe higher-order interaction vertices which will be generated during the RG flow even if they are not present at the UV scale. Further, we identify $y_{0} \equiv y_{\rmt}$ as the usual top Yukawa coupling as well as $\lambda_{h\psi,0} \equiv m_{\psi}$ and  $\lambda_{h\psi,1} \equiv \lambda_{h\psi}$ as the dimensionless dark fermion mass term and Higgs-portal coupling, respectively. 
We focus on the basic stabilization mechanisms of the Higgs potential, which take place in the SYM regime, i.e., for energy scales much higher than the electroweak scale.
In this regime, the simple approximation of the potentials in terms of polynomials is sufficient to analyze the system \cite{Borchardt:2016xju}.

The perturbative RG evolution of the quartic Higgs self-coupling already provides a good indicator for the occurrence of an instability if the impact of higher-dimensional couplings is neglected at UV scales. In this case, the zero crossing of the coupling determines the instability scale. 
This is corroborated by investigations of the full non-perturbative functional flow of the potential as well as polynomial approximations thereof \cite{Borchardt:2016xju,Gies:2017zwf}. The latter serve as the simplest possible extension of the perturbative flow equations.

Permitting the existence of higher-dimensional operators at the UV scale, the instability scale can be shifted towards larger scales. In this case, the metastability of the potential arises as an intricate interplay of the nontrivial bare structure of the Higgs potential at the UV cutoff scale and the fluctuating fields. 
A second or even several competing minima can occur. Then, polynomial truncations might no longer be an optimal choice for the proper tracking of the RG evolution of the Higgs potential. Instead, global information in field space is required to investigate all implications properly.

Nonetheless, polynomial approximations of the potential are still able to capture the relevant information of the occurrence of an instability for bare interactions of polynomial type \cite{Borchardt:2016xju}.  For this particular class, the radius of convergence of the potential in field space is finite and shrinks during the RG flow. Due to the fact that the second minimum usually occurs close to the cutoff scale, the stability issue can be addressed within these polynomial truncations.

The running of the various polynomial couplings can be extracted by suitable projections of the beta functionals of the potentials.
The flow equation for the quartic scalar coupling, $\pt \lambda_{2} = \pt u''(0)$ in the SYM regime, reads
\begin{align}
 \pt \lambda_{2} &= \frac{1}{16\pi^{2}} \left[ \frac{9\lambda_{2}^{2}}{(1+m_{\phi}^{2})^{3}} + 4\lambda_{2}y_{\rmt}^{2} - 4y_{\rmt}^{4} \right.  \notag \\
 &\qquad\qquad \left. + \frac{4\lamportal^{2}}{(1+m_{\psi}^{2})^2} - \frac{16 m_{\psi}^2\lamportal^{2}}{(1+m_{\psi}^{2})^3} \right]\,.
 \label{eq:quartic}
\end{align}
Here, we have neglected any term coming from higher-dimensional operators except for the Higgs-portal coupling, i.e., we have truncated the potentials at the simplest nontrivial order $(\Nu,\Nh,N_{\psi}) = (2,0,1)$. 
Further, we have neglected contributions from the anomalous dimensions within the threshold functions. These correspond to various resummed diagrams and thus higher-loop orders. 
As we have derived the flow equation from the Wetterich equation, threshold effects of massive modes are automatically incorporated. Due to the fact that the top quark is massless in the SYM regime, no threshold corrections appear for top fluctuations. By contrast, a mass term for the dark fermion $\psi$ is not forbidden by the symmetries of the action \eqref{eq:action}, thus modifying the running for nonvanishing masses.

First, we analyze the flow equation for the quartic coupling in the massless limit, $m_{\psi}\to 0$ and $m_{\phi}\to 0$, i.e., we set the threshold corrections to zero and obtain the one-loop coefficients in the deep Euclidean region. 
Crucially, the impact of the portal coupling $\lambda_{h\psi}$ comes with the opposite sign than the pure top quark loop within this limit. This might be rather surprising at first sight. It is commonly known that fermionic fluctuations cause the diminishing of the quartic coupling towards the UV, resulting in the stability constraints on the Higgs mass. Thus, the inclusion of further fermion species usually lowers the instability scale. However, the dark fermion couples to the scalar field in a different manner than the usual Yukawa-type structure as dictated by symmetry. 
Of course, a dark-fermion loop generally comes with a negative sign, just as the top loop does. This is due to their fermionic nature which can be seen in the flow equation for the scalar potential, cf.~Eq.~\eqref{eq:flowU}. But, we obtain a relative sign between the pure top-quark loop and the dark-fermion loop in the flow equation of the quartic coupling due to the different vertex structure of the different scalar-fermion interactions. Hence, the dimension-$5$ operator can effectively reduce the Higgs mass for a fixed cutoff scale. From another perspective, this implies that the instability scale is shifted towards larger scales.

However, this lowering mechanism is only present in a certain range of the dimensionless dark-fermion mass-parameter. Including threshold effects of the dark fermion, we can directly infer from the second line in Eq.~\eqref{eq:quartic} that the contribution of the dark fermion changes sign at $m_{\psi} = 1/3$, cf.~\cite{Eichhorn:2018vah}. Thus, as long as $m_{\psi} < 1/3$, the quartic coupling $\lambda_{2}$ is effectively reduced during the flow from UV to IR. In case $m_{\psi} > 1/3$, the dark fermion loops contribute to the growth of the Higgs mass in the IR, similar to the top quark. 
As the dimensionless parameter $m_{\psi}$ generically grows towards the IR, both effects will contribute to the running of the scalar sector if the initial parameter at the UV scale is below the critical value of $1/3$. 
Nonetheless, the latter effect contributes only over a comparatively short amount of RG ``time'' because the dark fermion decouples from the flow as soon as $m_{\psi} \gg 1$ due to threshold effects. 
Thus, we can divide the impact of the dark fermion on the running of the Higgs sector into three different regimes; stabilizing, destabilizing, and decoupling.

It depends on the precise initial parameters whether the stabilizing or destabilizing mechanism dominates the RG running of the scalar sector. 
In case the initial mass parameter is larger than one-third of the cutoff scale, i.e., $m_{\psi}(k=\Lambda)>1/3$, the Higgs mass increases due to the dark fermion. For $m_{\psi}(k=\Lambda)$ smaller but sufficiently close to $1/3$, the diminishing effect is present for a few RG steps but erased by the destabilizing effect for $m_{\psi}>1/3$ before the decoupling limit governs the flow (depending on the strength of the portal coupling as well as the scalar sector itself). Therefore, the Higgs mass is only reduced for sufficiently small bare $m_{\psi}(k=\Lambda)$ such that the contribution from the integrated running in the stabilizing regime overwhelms the contribution from the destabilizing regime.

In order to check in which parameter regions the dark fermion diminishes the lower mass bound or compounds the stability issue, we investigate in detail the RG evolution of the proposed toy model and perform various truncation tests to substantiate the picture arising from this simple analysis in the following.

\subsection{Functional analysis of the Higgs potential}

Solving the system of flow equations for the scalar, Yukawa, and portal potential as well as for the anomalous dimensions of the fields allows us to investigate the Higgs-stability problem on a global level in field space. However, this is a rather computing-time consuming endeavor as a system of coupled nonlinear partial differential equations has to be solved over many orders of magnitude with a sufficiently high precision to separate the UV cutoff scale $\Lambda_\text{EFT}$ from the electroweak scale. 
Therefore, we begin the investigation with the simplest possible extension of the perturbatively renormalizable flow equations by expanding the potentials in polynomials up to a finite order.

The beta functionals of the potentials can be viewed as closed-form expressions for the flow equations for arbitrary higher-order operators. 
The main advantage of the expansion of the potentials in Taylor series around the minimum of the scalar potential is that the system of partial differential equations is reduced to a system of ordinary differential equations. A typical flow for Higgs masses close to the lower bound starts in the SYM regime. We fine-tune the initial mass parameter for the scalar field such that fermionic fluctuations drive the system into the SSB regime as soon as the RG scale approaches the electroweak scale and the Higgs potential develops a nonvanishing minimum. At this point, we switch to expansions of the potentials which are given by Taylor series around the flowing minimum. The latter is the main point of interest in field space because it determines the mass and couplings of the particles. 
The nonvanishing vacuum expectation value of the scalar field induces a mass for the top quark as well as for the Higgs excitation. 
Therefore, all particles involved in the RG flow become massive and the flow freezes out completely. 
Further, we vary the bare Yukawa coupling $y_{0,\Lambda}$ to obtain a top mass of $173$ GeV. For the moment, we leave the Higgs mass as a free parameter which depends on the UV cutoff scale $\Lambda$, the bare quartic coupling $\lambda_{2,\Lambda}$, as well as other details in the bare action encoded in the higher-dimensional couplings $\lambda_{3,\Lambda}, \lambda_{4,\Lambda}, \dots$, $y_{1,\Lambda}, y_{2,\Lambda}, \dots$ and the dark sector $m_{\psi,\Lambda}, \lambda_{h\psi,1,\Lambda}, \lambda_{h\psi,2,\Lambda},\dots$ .

In order to check the convergence properties of the polynomial approximation of the full potential, we perform calculations for different truncation orders $\Nu$, $\Nh$, and $N_{\psi}$ and compare the extracted masses for the scalar field. Here we follow the same strategy as in \cite{Gies:2013fua,Borchardt:2016xju} for the model without dark fermions. 
We observe the following behavior. As most of the important physics is stored in the shape of the potentials at the Fermi scale, the expansion around the flowing minimum of the scalar potential $u$ is well suited to extract the masses of the particles and we obtain a rather fast convergence. With respect to the truncation order of the scalar potential, $\Nu = 2$ is the simplest nontrivial order. Increasing $\Nu$, i.e., including more and more higher-dimensional scalar self-interactions, we find $\Nu = 4$ as an optimal truncation parameter for fixed $\Nh$ and $N_{\psi}$. From $\Nu = 2$ to $\Nu = 4$ we find a slight deviation in the computed Higgs mass by a few percent in small as well as large coupling regimes. For $\Nu > 4$, we find no deviations from the results obtained for $\Nu = 4$ within our numerical accuracy. Performing the same investigation for fixed $\Nu$ and $N_{\psi}$ and varying $\Nh$, we find that the resulting Higgs masses are converged for $\Nh = 2$. Similarly, we obtain $N_{\psi} = 1$ as an optimal truncation order regarding convergence and computing time, thus demonstrating the remarkable convergence of the polynomial truncation for the present purpose. 

In order to test the convergence regarding the derivative expansion of the effective average action, we compare leading order to next-to-leading order results. We drop the running of the kinetic terms at leading order by setting the anomalous dimensions to zero in the flow equations of the potentials (local potential approximation). We find mild differences about $5{\,\%}$ for the lower mass bounds. This difference is mainly caused by the fact that the local potential approximation does not include all one-loop contributions given by one-particle reducible diagrams. These are stored in the anomalous dimensions which are not present inside a threshold function, i.e., the contribution of the anomalous dimensions which are separated in the first lines on the right-hand sides of Eqs.~\eqref{eq:flowU}-\eqref{eq:flowHpsi} which modify the canonical scaling dimensions. Including these contributions but setting the anomalous dimensions inside the threshold functions to zero, we obtain changes of merely a few $100$ MeV of the resulting Higgs masses compared to the leading-order result of the derivative expansion.

\subsection{Higgs mass bounds and phenomenological implications}

\begin{figure}[t!]
\centering
\includegraphics[width=0.98\columnwidth]{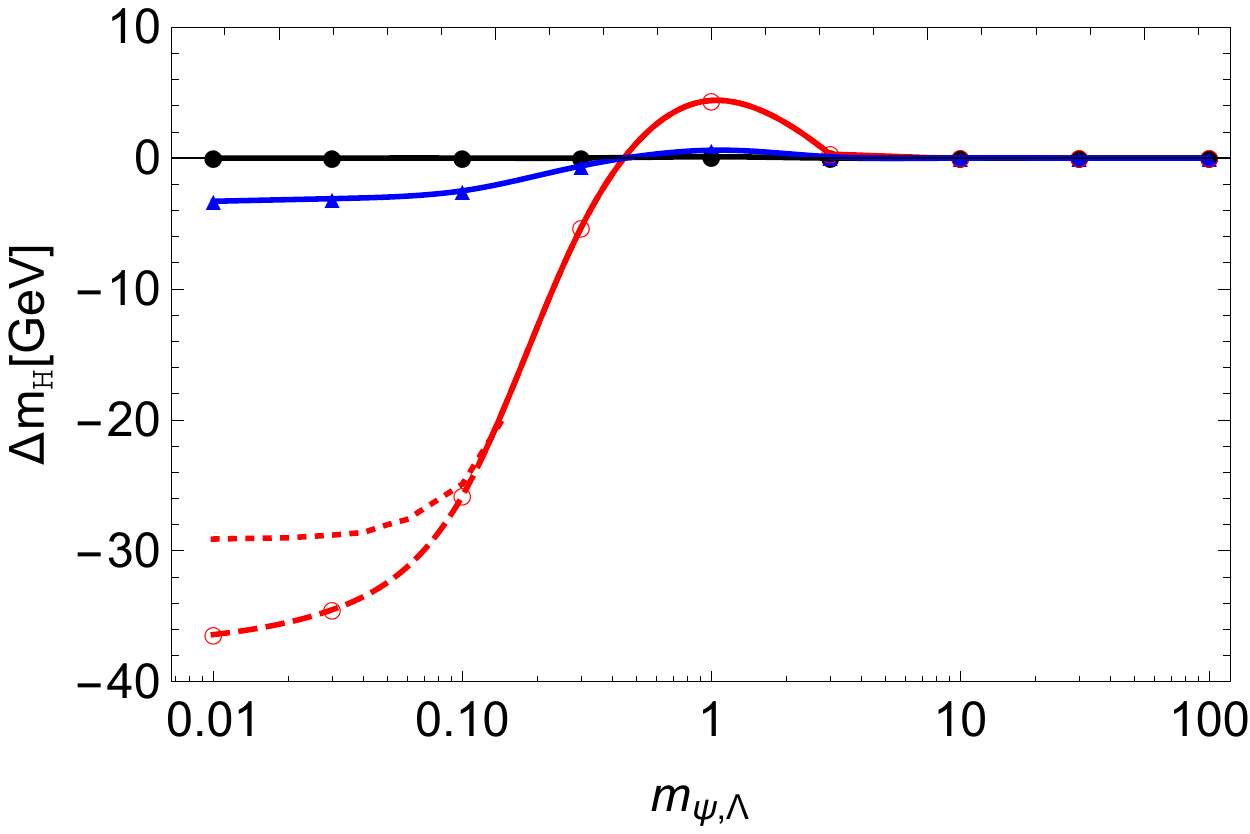}
\caption{Relative shift of the Higgs mass from the conventional lower mass bound as a function of the dimensionless dark fermion mass parameter at the cutoff scale for $\Lambda = 10^{5}$ GeV. The black curve with black filled circles denotes values for $\lambda_{h\psi,\Lambda} = 0.1$. The blue curve with triangles and the red curve with empty circles show results for $\lambda_{h\psi,\Lambda} = 1$ and $\lambda_{h\psi,\Lambda} = 3$, respectively. Solid lines indicate that the shift can be obtained without introducing a metastability in the effective Higgs potential. A second minimum is present in the effective potential for dashed curves. The dotted branch depicts a shift in the Higgs mass with stable potential which is stabilized by a nonvanishing scalar sector in the UV.
}
\label{fig:HiggsDifference}
\end{figure}

In Fig.~\ref{fig:HiggsDifference} we plot the shift of the low energy Higgs mass as a function of the dimensionless dark fermion mass parameter at the cutoff scale $\Lambda$ for different values of the dimensionless portal coupling $\lambda_{h\psi,\Lambda}$.\footnote{ 
{The fact that we use $m_{\psi,\Lambda}$ instead of the observable IR mass of the dark fermion is merely for reasons of convenience. As a rule of thumb, the dark fermion mass can be computed by the following simple relation $m_{\psi,\mathrm{IR}} = (m_{\psi,\Lambda}+ \frac{\lambda_{h\psi,\Lambda}}{64\pi^{2}} ) \Lambda $ which deviates from the actual dark fermion mass by at most $3\%$ for all tested parameter combinations. This approximation is based on the fact that the dark fermion sector renormalizes multiplicatively and is mainly dominated by its power counting behavior.}
} 
In this example, the cutoff scale is set to $\Lambda = 10^{5}$ GeV but similar conclusions hold also for other scales. The mass shift measures the departure from the conventional lower bound. For the considered toy model, the equivalent of the conventional lower bound is given by $m_{H,\mathrm{min}} = 94.9$ GeV for a cutoff scale $\Lambda = 10^{5}$ GeV. 
As in our preliminary analysis of the quartic coupling, we find that a diminishing of the lower mass bound is achieved for small bare mass parameters $m_{\psi,\Lambda}$. However, this effect is only measurable for sufficient strong portal couplings of $\mathcal{O}(1)$ or even larger. For $\lambda_{h\psi,\Lambda} = 0.1$, depicted as a black line with filled circles, the deviation from the conventional lower mass bound without dark sector and vanishing higher-order operators at the cutoff scale is less than $0.1$ GeV. 
The effect of the dark fermions diminishing the lower Higgs-mass bound grows rapidly for larger values of $\lambda_{h\psi,\Lambda}$. We depict the results for $\lambda_{h\psi,\Lambda} = 1$ and $\lambda_{h\psi,\Lambda} = 3$ as a blue curve with triangles and a red curve with empty circles, respectively. The largest mass shift is achieved for a vanishing dark fermion mass parameter at the cutoff scale. Already for $m_{\psi,\Lambda} < 0.01$ we observe a saturation of the curve towards the limit $m_{\psi,\Lambda} \to 0$. 

By contrast, neither in the small nor large coupling regime a deviation from the conventional lower mass bound can be observed for large bare mass parameters. Of course, this is expected as the threshold effects of the dark fermion dominate the impact on the RG flow of the scalar sector already in the UV in this limit. This can be circumvented for extremely large portal couplings, $\lambda_{h\psi} \gtrsim m_{\psi}^{2}$, such that the decoupling in the flow equation of the Higgs potential is compensated by the large coupling. However, the dark fermion fluctuations induce larger Higgs masses, i.e., lower the instability scale, in this limit as can be directly inferred from Eq.~\eqref{eq:quartic}. Therefore, we do not further explore the parameter space in this regime.

In the intermediate mass regime, we observe the previously discussed change in the effect of dark-fermion fluctuations on the lower Higgs-mass bound. Instead of diminishing the lower mass bound, the fermionic fluctuations start raising it close to the critical value of $m_{\psi} = 1/3$. At this critical value, the dark-fermion contribution to the running of the quartic coupling changes sign. The raising of the lower Higgs-mass bound approaches a maximum around $m_{\psi,\Lambda} \approx 1$, cf.~Fig.~\ref{fig:HiggsDifference}. We observed that a large cutoff portal-coupling $\lambda_{\psi,\Lambda}$ quickly enhances the lowering effect for $m_{\psi,\Lambda} < 1/3$. Similarly, the effect of raising the Higgs mass bound for $m_{\psi,\Lambda} \approx 1$ quickly increases with growing $\lambda_{\psi,\Lambda}$.

Depending on the specific values of the dark-fermion mass and the portal coupling, dark fermions have the potential to lower the Higgs mass by an arbitrary amount as can be seen in Fig.~\ref{fig:HiggsDifference}. Nevertheless, this does not mean that the Higgs potential remains stable during the RG flow. We observe that the dark fermions can indeed stabilize the Higgs potential and thus raise the instability scale but only for a certain finite amount. Once a critical upper value of the shift in the Higgs mass is passed, the effective average potential starts to develop a second nontrivial minimum at large field values usually below but close to the cutoff scale $\Lambda$, rendering the Higgs potential metastable.

This is similar to investigations on the impact of higher-dimensional operators of just standard-model degrees of freedom. Within these examples, the instability scale can also be shifted towards larger scales. As the running of higher-dimensional couplings is mainly governed by their power-counting dimension, they quickly die out. This implies that they can only stabilize the Higgs potential by $1$-$2$ orders of magnitude if the dimensionless higher-order couplings are of order one.
We observe a similar effect in the model with dark fermions where the stabilizing effect is also caused by a power-counting nonrenormalizable operator. By contrast, the instability scale can easily be shifted towards larger scales for scalar dark matter, where the portal coupling is marginal \cite{Silveira:1985rk,McDonald:1993ex,Burgess:2000yq,Bento:2001yk,Patt:2006fw,Eichhorn:2014qka}. 

In Fig.~\ref{fig:HiggsDifference} we indicate this fact by different codings of the interpolating lines. Solid lines depict shifts in the Higgs mass for which the scalar potential $u$ is stable during the entire RG flow. 
Dashed lines depict a possible diminishing of the lower mass bound accompanied by a second minimum of the effective potential beyond the Fermi scale. In almost all cases this is the global one, thereby again rendering the potential metastable. This occurs only for sufficiently strong portal couplings and small dark-fermion masses. Then the RG flow of the quartic coupling is driven towards negative values. These are not compensated for by positive higher-dimensional couplings of the scalar sector to render the potential stable. However, this effect can be mitigated if a positive and sufficient large quartic or higher-dimensional coupling in the scalar sector is present in the UV. The lowest possible value of the Higgs mass with stable Higgs potential is indicated as a dotted line for $\lambda_{h\psi,\Lambda} =3$. For this, we scanned the parameter region by allowing for such positive bare couplings and examined the transition from a stable to an unstable potential. Thus, any shift of the Higgs mass above the dotted or solid curve can be reached for $\lambda_{h\psi,\Lambda} =3$ without introducing a second minimum in the effective potential.

\begin{figure}[t!]
\centering
\includegraphics[width=0.98\columnwidth]{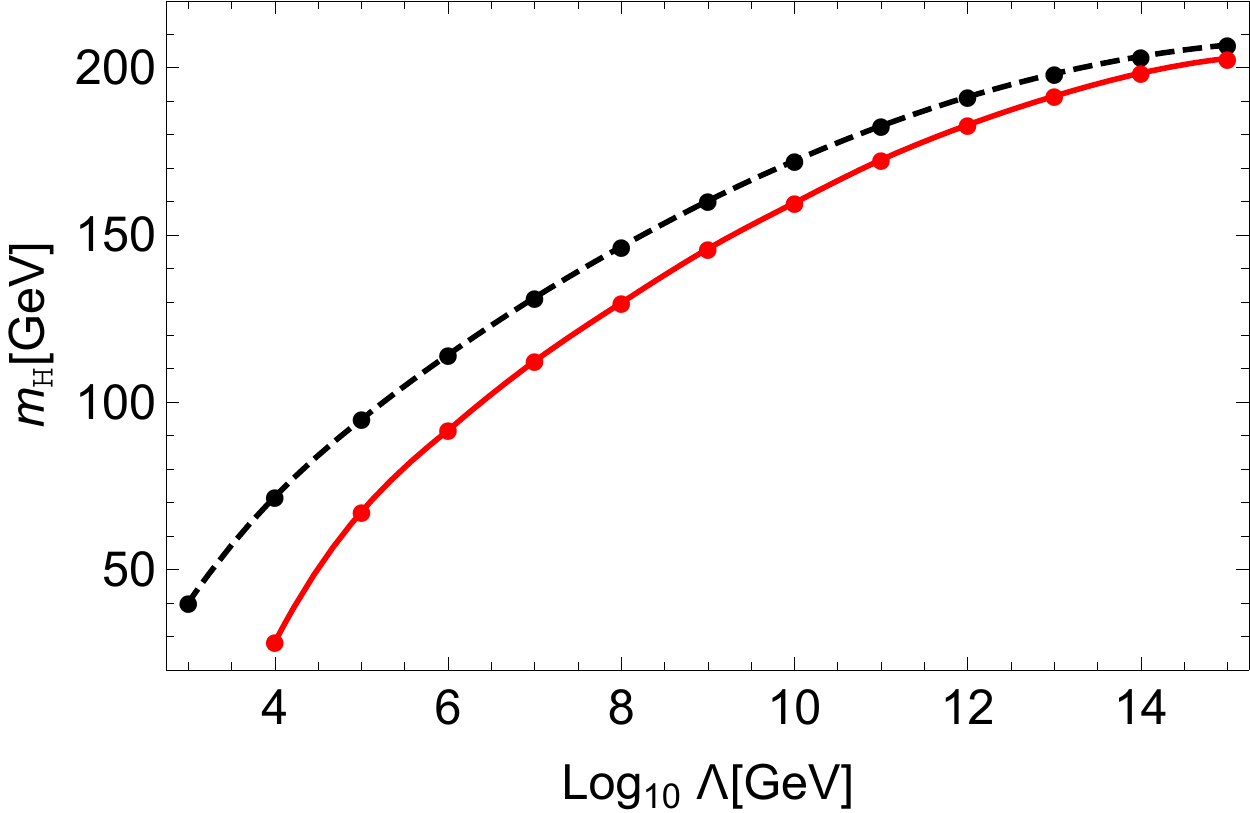}
\caption{Lower Higgs-mass bounds with a single minimum in the effective Higgs potential. The black-dotted line depicts the conventional lower mass bound for the considered toy model without dark matter or the influence of any higher-dimensional operator in the UV. The red solid line demonstrates how fermionic dark matter, as well as other higher-dimensional operators, can diminish this bound without introducing a metastability in the effective potential. Thus the cutoff scale of new physics $\Lambda$ is increased.}
\label{fig:HiggsMassBounds}
\end{figure}

In Fig.~\ref{fig:HiggsMassBounds}, we plot the conventional lower Higgs mass bound for this toy model, i.e., all higher-dimensional couplings are zero at the cutoff scale, as a black dashed line for various cutoff scales $\Lambda$. In addition, the red curve denotes the diminishing of the lower bound due to the portal coupling to dark fermions for fully stable Higgs potentials during the entire RG flow for bare couplings of $\mathcal{O}(1)$. A further diminishing would result in a metastable potential due to the complex interplay between the bare structure of the potentials and the quantum fluctuations. Similarly to investigations in models without a dark fermion, we find that the stability scale can be shifted by roughly one order of magnitude towards larger cutoff scales by higher-dimensional operators.

Finally, we address the question of the validity of the polynomial approximation of the potential to investigate the stability problem globally in field space. To that end, we solve the full partial differential equation for the scalar potential $u$ in leading order in the derivative expansion. In order to minimize the computational effort, we treat the couplings of the other sectors as external sources, i.e., we neglect their RG running forced by quantum fluctuations. Of course, this does not capture the mutual back reactions between the different sectors but is sufficient to investigate the occurrence of a possible second minimum on a qualitative level. We find that the polynomial truncation of the potential is able to describe the occurrence of a second minimum during the flow as in the previous studies without a dark fermion \cite{Borchardt:2016xju}. The radius of convergence is large enough to appropriately track the development of a second minimum at large scales during the RG flow. For instance, for $\Lambda = 10^{5}$ GeV, we have examined that both, the full flow as well as the polynomial approximation of the potential, allow for the same shift of ${\approx} 24$ GeV of the lower mass bound due to the portal coupling with a stable electroweak minimum. In both approximations, increasing the portal coupling, which results in a further decrease of the Higgs mass, causes the occurrence of a second minimum.

The full flow of the potential shows the same behavior as investigated in previous toy models without fermionic dark matter. Thus, the zero crossing of the quartic coupling provides a good estimate for the instability scale as long as a further stabilization by higher-dimensional operators is absent.
However, the latter does not alter the order of magnitude of the shift of the instability scale but causes small quantitative modifications towards larger scales. In the following, we will use these results of the simple toy model to address the implications for the standard-model Higgs potential.  In particular, we will confront our results with experimental constraints on the relic abundance which further constrains the parameters of the dark sector.

\section{Fermionic Dark Matter and the Standard-Model Higgs-potential}
\label{sec:Pheno}

To compare our analysis with experimental bounds on the fermionic dark matter couplings, we supplement the simple model with the standard model gauge structure, cf.~\cite{Eichhorn:2015kea}, without explicitly gauging the Higgs. In particular, we add an $SU(3)$ gauge group under which we charge the $n_f$ Yukawa-type fermions, but neither the dark-matter fermions nor the Higgs, modifying the corresponding effective action, cf.~Eq.~\eqref{eq:action}, to read
\begin{align}
\label{eq:eff_action_toySM}
 \Gamma_{k} = \int_{x} \bigg[ 
     &\frac{Z_G}{4}F_{\mu\nu}^a\,F^{a\,\mu\nu}
     +\frac{Z_{\phi}}{2} \partial_{\mu}\phi \partial^{\mu}\phi 
     + \sum_{i}^{n_f} Z_{i}\, \bar{q}_i \I \slashed{D} q_i
     \notag\\
     &+ Z_{\psi}\, \psib \I \slashed{\partial}\psi  
     + U(\rho) 
     + \frac{\I}{\sqrt{2}} H_{\mathrm{t}}(\rho) \phi\, \bar{q}_tq_t  
     \notag\\
     &+ \I H_{\psi}(\rho) \psib\psi 
 \bigg].
\end{align}
We also keep neglecting all Yukawa couplings from the non-heavy fermions $q_{i\neq t}$ because of their negligible impact on the running of the Higgs potential. This captures the influence of the strong interaction in the standard model above $\Lambda_\text{QCD}$.

Gauging the scalar under the electroweak gauge-group, i.e., transitioning to a Higgs-doublet, would require our model to capture the full electroweak sector of the standard model to observe the freeze-out of all massive modes correctly. 
Instead, we add a fiducial gauge-contribution to the running of the dimensionless effective scalar potential $u(\rho)$ and the Yukawa potential $h_\rmt(\rho)$. It has been checked that such a model captures all of the standard-model structure required for a quantitative discussion of the lower Higgs-mass bound \cite{Eichhorn:2015kea}, but avoids intricate questions arising from the involved Higgs-gauge boson interplay \cite{tHooft:1979hnm,Osterwalder:1977pc,Banks:1979fi,Frohlich:1980gj,Frohlich:1981yi,Maas:2012tj,Maas:2013aia,Maas:2017wzi}. The dark fermions are left uncharged under both the strong and the fiducial electroweak gauge sector.

The running of this toy standard model is captured by adding the following additional contributions from fiducial electroweak gauge loops as well as the strong sector to Eqs~\eqref{eq:flowU}-\eqref{eq:flowHpsi}:
\begin{align}
    \partial_t u(\tilde{\rho})\Big|_\text{toy-SM} &= 
    \partial_t u(\tilde{\rho}) 
    +\frac{4v_{d}}{d} \frac{c_u}{1+\frac{\tilde\rho\,g_F^2}{2}},
    \\
    \partial_t h_\rmt(\tilde{\rho})\Big|_\text{toy-SM} &=
    \partial_t h_\rmt(\tilde{\rho})
    + 2\,v_d\,c_\rmt\,g_F^2 h_\rmt
    \\\notag
    &\hspace*{-25pt}
    - 4\,v_d\frac{N_c^2-1}{2N_c}g_s^2\,h_\rmt(3-\xi)\,
     l_{1,1}^{(\rmB\rmF)d}(0,\omega_{\rmt}; \eta_{G},\eta_{\rmt}).
\end{align}
The strong gauge coupling is denoted by $g_{s}$. The terms containing the fiducial gauge coupling $g_{F}$ model electroweak contributions. In all following numerical computations we will employ Landau gauge, i.e., $\xi=0$, which is also a fixed point of the RG flow \cite{Ellwanger:1995qf}. Further, one has to account for closed fermionic loops in the scalar potential and anomalous dimension, that now contribute with color multiplicity, i.e., $d_\gamma\rightarrow d_\gamma\,N_c$ in the second line of Eqs~\eqref{eq:flowU} and \eqref{eq:etaPhi}. Finally, there are also additional gauge contributions to the fermionic anomalous dimension, cf.~Eq.~(C9) in~\cite{Eichhorn:2015kea}, but we refrain from showing them here since they cancel out in Landau gauge ($\xi=0$). With $g_{F} = 0.57$, $c_\rmt=\frac{97}{30}$, and $c_u=9$ the running of the conventional top-Yukawa coupling $y_\rmt = h_\rmt(0)$ and the quartic Higgs coupling $\lambda_{2} = u''(0)$ falls back to the known standard-model 1-loop running when threshold effects are neglected \cite{Eichhorn:2015kea}.

\subsection{Parameter space of fermionic dark matter}

As we have carefully analyzed in Sec.~\ref{sec:FlowEquations}, the lowest non-trivial truncation of the polynomial approximation, i.e., $(N_u,N_h,N_\psi) = (2,0,1)$, already captures the qualitative features of the effects of fermionic dark matter on the lower Higgs-mass bound. We will therefore present results in the $(2,0,1)$-scheme and confirm the validity of this approximation at specific points, cf.~App.~\ref{app:appendix}. We fix the coupling values of the strong gauge, the top Yukawa and the scalar quartic coupling to match the standard model values $g_s(1\rm TeV)=1.060$, $y_\rmt(1\rm TeV) = 0.867$ and $\lambda_{2}(1\rm TeV)=0.202$ at $1$ TeV at one loop. Then we map out the possible parameter space of fermionic dark matter mass $m_\psi(1\rm TeV)$ and portal coupling $\lambda_{h\psi}(1\rm TeV)$ at the matching scale. 
Since we only consider dark-matter masses above $\bar{m}_\psi>1\,\rm TeV$, the dark matter couplings will always freeze out above $1\,\rm TeV$. Below $1\,\rm TeV$ they scale near-canonically. 
Hence, the dimensionful values of $\bar{m}_\psi$ and $\bar{\lambda}_{h\psi}$ essentially do not run below $1\rm TeV$. We confirm the validity of this approximation in App.~\ref{app:appendix}.

\begin{figure}[t!]
\centering
\includegraphics[width=\columnwidth]{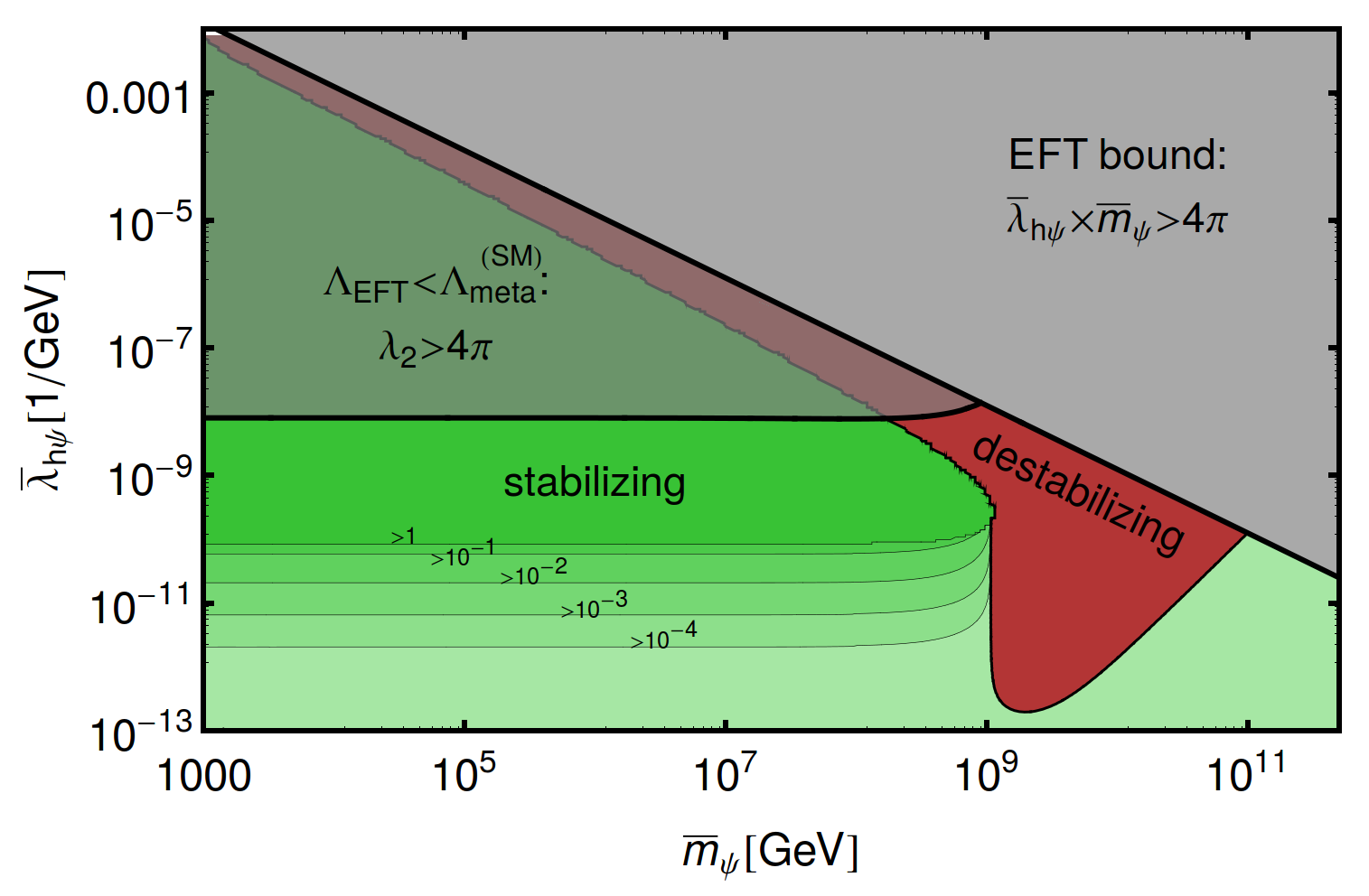}
\caption{
Regions in the parameter space of dark matter mass $\bar{m}_\psi$ and portal coupling $\bar\lambda_{h\psi}$, in which the Higgs potential is (de)stabilized. In the green (red) region the Higgs stability-bound is shifted to larger (smaller) values. The thin contours show the relative increase of the metastability scale, i.e., $(\Lambda_\mathrm{meta}^{(\mathrm{fdmSM})}-\Lambda_\mathrm{meta}^{(\mathrm{SM})})/\Lambda_\mathrm{meta}^{(\mathrm{SM})}$. The shaded regions indicate where a perturbative EFT description, either in the fermionic sector or in the scalar sector breaks down.}
\label{fig:stab-destab}
\end{figure}

Fig.~\ref{fig:stab-destab} shows the difference of the metastability scale with fermionic dark matter and the standard-model metastability scale, i.e., $\Lambda_\mathrm{meta}^{(\mathrm{ fdmSM})}-\Lambda_\mathrm{meta}^{(\mathrm{SM})}$. In the simple $(2,0,1)$-scheme, the metastability scale is estimated by the respective scale at which the quartic coupling turns negative. For our toy standard-model (without fermionic dark matter) this amounts to $\Lambda_\mathrm{meta}^{(\mathrm{SM})}=2.54\times10^{9}$ GeV. The contours within the stabilizing (green) regime indicate by how much the metastability scale can be increased.
In accordance with the analysis of the simple Higgs-Yukawa model in Sec.~\ref{sec:FlowEquations}, we again find a destabilizing regime whenever $\bar{m}_\psi\approx\Lambda_\mathrm{meta}^{(\mathrm{SM})}$, cf.~red region in Fig.~\ref{fig:stab-destab}.

The functional RG is a non-perturbative tool. Hence our analysis might still capture the correct dynamics in the non-EFT regime. Nevertheless, we also determine the scale, $\Lambda_\text{EFT}^\text{( fdmSM)}$,  at which the perturbative EFT conditions $\bar{\lambda}_{h\psi}\bar{m}_\psi<4\pi$ and $\lambda_2<4\pi$ break down. While for the standard model on its own this trans-Planckian scale is as high as $\Lambda_\text{EFT}^\text{(SM)}\approx 10^{41}\,\rm GeV$, the influence of the fermionic dark sector can significantly lower this bound. We mark these regions in Fig.~\ref{fig:stab-destab} as dark-gray shaded. The upper right triangle in parameter space shows where the EFT criterion $\bar{\lambda}_{h\psi}\bar{m}_\psi<4\pi$ is already violated at 1 TeV. We do not perform further computations in this regime. The region on the upper left-hand side $\lambda_2>4\pi$ shows where the Higgs-quartic coupling leaves the EFT-regime below the standard metastability scale, i.e., $\Lambda_\text{EFT}^\text{( fdmSM)} < \Lambda_\text{meta}^\text{(SM)}$. Shortly thereafter, it also runs into a perturbative Landau pole. This occurs due to too large screening contributions from the fermionic dark matter. While the fermionic dark sector remains of stabilizing nature, it nevertheless fails to increase the scale of new physics $\Lambda_\text{new-phys}^\text{( fdmSM)} = \text{Min}(\Lambda_\text{meta}^\text{( fdmSM)},\Lambda_\text{EFT}^\text{(fdmSM)})$.

Finally, we check the results obtained in $(N_u,N_h,N_\psi) = (2,0,1)$ against the ones obtained in the optimal truncation order $(N_u,N_h,N_\psi) = (4,2,1)$, cf.~Sec.~\ref{sec:FlowEquations}. This requires a tuning of the bare mass parameter for each point individually. For all points that we tested the qualitative behavior agrees with the lower order truncation, with quantitative shifts of the contours up to $9\%$ at log-log scales.
Details on this analysis can be found in App.~\ref{app:appendix}.

\subsection{Cosmologically viable fermionic dark matter requires additional new physics}
The standard-model analysis confirms the findings of the simple Higgs-Yukawa model. Fermionic dark matter can increase the new physics scale $\Lambda_\text{new-phys}^\text{(fdmSM)}>\Lambda_\text{new-phys}^\text{(SM)}$ to which the standard model can be extrapolated as a stable EFT. For this, the dark-fermion mass has to be light in comparison to the metastability scale, in order to avoid a complete decoupling of the dark sector or a Yukawa-like influence on the running of the Higgs-potential. Secondly, the portal coupling has to be very weak, i.e., $\bar{\lambda}_{h\psi}<1/\Lambda_\mathrm{meta}^{(\mathrm{SM})}$. Otherwise, it induces a breakdown of the EFT in the form of a Landau-pole-like instability below $\Lambda_\mathrm{meta}^{(\mathrm{SM})}$. Even if both requirements are met, the Higgs-portal coupling is only able to increase $\Lambda_\text{new-phys}$ by about one order of magnitude, cf.~Fig.~\ref{fig:HiggsMassBounds}. Therefore, in contrast to scalar dark matter, fermionic dark matter cannot extend the standard model as a stable EFT up to the Planck scale, even within the present non-perturbative analysis.

Furthermore, the regime in which fermionic dark matter has a slight stabilizing effect, i.e., $\bar{\lambda}_{h\psi}<10^{-8}$ GeV$^{-1}$, is in direct conflict with relic density constraints. Perturbative studies \cite{Beniwal:2015sdl} suggest that any fermionic portal coupling $\bar{\lambda}_{h\psi}<10^{-3}$ GeV$^{-1}$ would oversaturate the relic-density bound and therefore overclose the universe.
Under the assumptions that fermionic dark matter is produced via a post-inflationary thermal freeze-out and the relic-density analysis is not drastically modified by non-perturbative effects, this observational constraint excludes the possibility of fermionic dark matter to stabilize the Higgs potential without introducing a Landau-pole-like instability below $\Lambda_\mathrm{meta}^{(\mathrm{SM})}$.

Therefore, if such fermionic dark matter was to be discovered experimentally, it would significantly impact the Higgs potential. The stabilizing fermionic fluctuations would drive the quartic coupling towards a perturbative Landau pole far below the current metastability scale, i.e., $\Lambda_\text{new-phys}^\text{( fdmSM)}\approx 1/\overline{\lambda}_{h\psi}$. In the present functional RG study, we find this behavior to persist at the non-perturbative level. Our study therefore tentatively suggests that fermionic dark matter, in agreement with relic-density constraints, either requires the transition into a regime of fully non-perturbative dynamics or suggests a larger dark sector close to $\Lambda_\text{new-phys}^\text{( fdmSM)}\approx 1/\bar{\lambda}_{h\psi}$ in which the Higgs-portal coupling appears as a higher-order interaction.

One possibility for a strongly coupled regime without the requirement of additional new degrees of freedom could be offered by a non-perturbative fixed point, as suggested by the analysis in \cite{Eichhorn:2018vah}. Whether tentative hints for the fixed point can be confirmed in extended studies and it might then be used to define a strongly coupled Higgs-portal sector remains a promising future study. This presumably requires the inclusion of higher-order operators in the fermionic sector.

\section{Conclusions}

We have investigated the global stability properties of the standard-model Higgs-potential with the functional renormalization group under the impact of fermionic dark matter coupled via a Higgs portal. The motivation for this study stems from the low-truncation (perturbative) indications that the fermionic portal coupling, contrary to Yukawa couplings, can reduce the lower Higgs-mass bound, i.e., delay the onset of the standard-model metastability. The reason for this distinct behavior is encoded in the different symmetry structure of the fermions coupled to the Higgs via Yukawa or portal interactions.

While the non-perturbative analysis confirms that the metastability scale can be shifted towards higher scales, the dimensionful nature of the portal coupling only allows stabilizing the Higgs-potential for about one order of magnitude. This already follows from a dimensional argument, as the Higgs portal to fermionic dark matter is a dimension-5 operator. We find that this perturbatively intuitive conclusion also holds at the non-perturbative level.
We numerically solve the flow equation for the scalar potential, as obtained from the functional RG equation at lowest order in the derivative expansion and confirm that the polynomial truncation sufficiently captures the structure of a possibly arising minimum.
Every attempt to stabilize the quartic coupling by significantly more than one order of magnitude leads to the emergence of a novel minimum for polynomial interactions. This resembles a low-scale Landau-pole like instability found in perturbative approximations. 
In non-perturbative truncations, the running couplings become singular in a similar fashion. 
This minimum quickly becomes the global minimum and the corresponding instability scale drops below the usual standard-model metastability scale -- hence invalidating the original motivation. We conclude that the dimension-5 portal coupling cannot significantly stabilize the Higgs-potential. This is due to its power-counting dominated suppression towards the IR in the vicinity of the Gau{\ss}ian fixed point. The same behavior was already found in the previous analysis of higher-order operators in the scalar and Yukawa potential \cite{Borchardt:2016xju,Gies:2017zwf}.
\\

Further, we supplement the simple Higgs-Yukawa model with an approximation of the gauge structure of the standard model, for which it was confirmed in \cite{Eichhorn:2015kea} that the running of couplings is quantitatively very close to that of the full standard model. We explicitly verify that all the above statements carry over to this gauge-supplemented model. This allows us to quantitatively analyze the influence of fermionic dark matter on the Higgs metastability-bound.
We find that the regime in which the Higgs potential can at least be stabilized by one order of magnitude relies on a very small portal coupling $\overline{\lambda}_{h\psi}<10^{-8}\,\text{GeV}^{-1}$. 
This is in conflict with the perturbative relic-density constraint $\overline{\lambda}_{h\psi}>10^{-3}\,\text{GeV}^{-1}$ for dark fermion masses $>100$ GeV obtained in, e.g. \cite{Beniwal:2015sdl}.

In turn, our study suggests that if fermionic dark matter obeying the relic-density constraint should be detected through its portal interaction with the Higgs, it would quickly lead to an increase of the scalar self-interactions encoded in the Higgs potential. 
This would drive the quartic coupling out of the perturbative regime. In this case, we find the formation of Landau-pole type structures in the couplings within our nonperturbative calculations.
The presence of fermionic dark matter thus requires additional new physics at scales no higher than about $\Lambda_\text{new-phys}^\text{(fdmSM)}\approx 1/\bar{\lambda}_{h\psi}$.
This is particularly intriguing because it suggests that fermionic dark matter either leads to a novel non-perturbative scaling regime, such as the asymptotically safe fixed-point proposal in \cite{Eichhorn:2018vah}, or requires an extended dark sector with additional degrees of freedom.

\textit{Acknowledgements:} 
We are grateful to A. Eichhorn and H. Gies for fruitful discussions and comments.
A.~Held acknowledges support by the Studienstiftung des deutschen Volkes as well as by the Emmy-Noether grant of the DFG awarded to Astrid Eichhorn under grant no. EI/1037-1. A.~Held also acknowledges the hospitality of Perimeter Institute for Theoretical Physics during the final stages of this work. Research at Perimeter Institute is supported by the Government of Canada through the Department of Innovation, Science, and Economic Development, and by the Province of Ontario through the Ministry of Research and Innovation. RS acknowledges support by a postdoc fellowship of the Carl-Zeiss foundation as well as by the DFG under Grants No. Gi328/9-1.

\appendix

\section{Comparison with the optimized truncation order}
\label{app:appendix}

In this appendix, we address the reliability of the lowest non-trivial truncation of the polynomial approximation, $(N_u,N_h,N_\psi) = (2,0,1)$, regarding the parameter space of fermionic dark matter of Sec.~\ref{sec:Pheno}. As discussed in detail for the simple toy model of Sec.~\ref{sec:FlowEquations}, the scheme $(N_u,N_h,N_\psi) = (4,2,1)$ captures all relevant informations of the polynomial flows regarding the IR masses of the particles. We observe the same behavior in the model with gauge boson contributions as well. Due to slight corrections of the higher-dimensional operators on the running of the quartic coupling, we obtain the same qualitative results but minor quantitative corrections on logarithmic scales. For instance, the instability scale $\Lambda_\mathrm{meta}^{(\mathrm{SM})}$ decreases as the quartic coupling is slightly driven faster towards zero mainly by the higher-dimensional operators of the Yukawa sector. As the quartic coupling runs logarithmically this causes a diminishing of the metastability scale by less than an order of magnitude. More precisely, we have $\log_{10}\Lambda_\mathrm{meta}^{(\mathrm{SM}),(4,2,1)}/\log_{10}\Lambda_\mathrm{meta}^{(\mathrm{SM}),(2,0,1)} = 0.91$. 

As the basic stabilizing mechanisms take place in the UV near the cutoff scale of the model, we get the same order of magnitude of the relative increase $(\Lambda_\mathrm{meta}^{(\mathrm{fdmSM})}-\Lambda_\mathrm{meta}^{(\mathrm{SM})})/\Lambda_\mathrm{meta}^{(\mathrm{SM})}$, if we choose the same UV initial conditions in the dark fermion sector for the different truncation schemes. However, due to the difference in the reference scale $\Lambda_\mathrm{meta}^{(\mathrm{SM})}$, we obtain quantitative corrections in the IR physics. This manifests as a shift of the contours towards smaller masses and larger portal couplings in the log-log graph of Fig.~\ref{fig:stab-destab}. This shift is almost universal for all test points in the plot. As a rule of thumb, the contours are shifted by $\approx -0.87$ in the horizontal direction (dark fermion mass) and $\approx 0.89$ in the vertical direction (portal coupling) on the presented log-log scale, i.e., by less than one order of magnitude. These results are stable with respect to a further increase of the truncation orders $(N_u,N_h,N_\psi)$.

Thus, we are still confronted with the fact that the relic-density constraint excludes the possibility of fermionic dark matter to stabilize the Higgs potential without introducing a Landau-pole-like instability below $\Lambda_\mathrm{meta}^{(\mathrm{SM})}$ for the advanced analysis. Therefore, we refrain from performing the analysis of Sec.~\ref{sec:Pheno} for the computationally expensive advanced truncation as the phenomenological interesting region in parameter space is still out of reach and the same conclusions remain.

\bibliography{Bibliography}  
\end{document}